\makeatletter
\def\input@path{{./tex/}{./bst/}}
\makeatother
\documentclass[%
reprint,
superscriptaddress,
 amsmath,amssymb,
 aps,
 prl,
floatfix,
]{revtex4-2}

\usepackage{graphicx}
\usepackage{dcolumn}
\usepackage{bm}
\usepackage{hyperref}
\usepackage{siunitx}
\usepackage{physics}
\usepackage[capitalise]{cleveref}
\usepackage[dvipsnames]{xcolor}  
\usepackage{ragged2e}

\usepackage[bf,justification=justified]{caption}
\usepackage{subcaption}
\newcommand{\labelphantom}[1]{%
	\parbox{0pt}{\phantomsubcaption\label{#1}}%
}

\newcommand{\cx}{|c_\mathrm{x}|^2}
\newcommand{\Upp}{U_\mathrm{pp}}
\newcommand{\gpp}{g_\mathrm{pp}}
\newcommand{\gtwo}{g^\mathrm{(2)}}
\newcommand{\Gammap}{\Gamma_\mathrm{p}}

\newcommand{\abbrev}[3]{
  \newcounter{#3}
  \setcounter{#3}{0}
  \newcommand{#1}{\ifnum\value{#3}<1{#2 (#3)}\else{#3}\fi\stepcounter{#3}}}

\abbrev{\DBR}{distributed Bragg reflector}{DBR}
\abbrev{\QW}{quantum well}{QW}
\abbrev{\sspd}{superconducting nanowire single--photon detectors}{SSPD}
\abbrev{\hbt}{Hanbury Brown and Twiss}{HBT}

\begin{document}


\title{Quantum correlations and dissipative blockade of polaritons in a tunable fiber cavity}

\author{Gian-Marco Schnüriger}
\email{gm.schnueriger@uni-bonn.de}
\affiliation{Institut f\"ur Quantenelektronik, ETH Z\"urich, 8093 Z\"urich, Switzerland.}
\affiliation{Physikalisches Institut, Universit\"at Bonn, 53115 Bonn, Germany.}
\author{Martin Kroner}%
\affiliation{Institut f\"ur Quantenelektronik, ETH Z\"urich, 8093 Z\"urich, Switzerland.}
\author{Emre Togan}%
\affiliation{Institut f\"ur Quantenelektronik, ETH Z\"urich, 8093 Z\"urich, Switzerland.}
\author{Patrick Knüppel}%
\affiliation{Department of Physics, Universit\"at Basel, 4056 Basel, Switzerland.}
\author{Aymeric Delteil}%
\affiliation{Univerit\'{e} Paris-Saclay, UVSQ, CNRS, GEMaC, 78000 Versailles, France.}
\author{Stefan F\"alt}
\affiliation{Laboratorium für Festk\"orperphysik, ETH Z\"urich, 8093 Z\"urich, Switzerland.}
\author{Werner Wegscheider}
\affiliation{Laboratorium für Festk\"orperphysik, ETH Z\"urich, 8093 Z\"urich, Switzerland.}
\author{Atac Imamoglu}
\affiliation{Institut f\"ur Quantenelektronik, ETH Z\"urich, 8093 Z\"urich, Switzerland.}

\date{\today}

\begin{abstract} 
Cavity exciton--polaritons are quasiparticles that form when quantum well excitons hybridize with a cavity mode \cite{Carusotto2013}. Here, we carry out photon correlation measurements under continuous wave resonant laser excitation to demonstrate quantum correlations between cavity--polaritons. Our experiments reveal an unexpectedly strong dependence of polariton interactions on cavity--exciton detuning. When the polaritons are predominantly exciton-like, we observe a transition from photon antibunching to bunching as the laser is tuned across the polariton resonance, in agreement with a simple Kerr-nonlinearity model \cite{Verger2006,Ferretti2012}. When the lower-branch polariton energy is tuned to induce a two-polariton Feshbach resonance with the biexciton mode, the degree of polariton antibunching becomes independent of the laser detuning: we explain our finding by invoking a dissipative blockade mechanism arising from large biexciton broadening. Our experiments demonstrate that the strong polariton blockade regime would be achieved by reducing the polariton decay rate by a factor of 10.
\end{abstract}

\maketitle


The elementary optical excitations of a semiconductor quantum well that is embedded inside a microcavity are exciton--polaritons. Unlike cavity photons, these quasiparticles exhibit finite Kerr-type interactions stemming from their exciton character \cite{Ciuti1998}. Concurrently, low-momentum polaritons are immune to disorder scattering due to their ultra-light mass, determined by the dispersion of the confined cavity mode. The combination of these features has made it possible to realize several ground-breaking experiments, ranging from polariton condensation \cite{Kasprzak2006} to the exploration of Kardar-Parisi-Zhang physics \cite{Fontaine2022}. All of the physics studied to date allowed for a description based on a mean-field treatment, since the strength of the nonlinearity at the single-polariton level was much weaker than the dissipation rate. It is widely accepted that reaching the limit where quantum correlations between single polaritons become prominent would initiate a new field of strongly correlated photons \cite{Verger2006,Imamoglu1997}.

In general, the observation of quantum correlations requires the interaction energy $\Upp = \gpp/A \simeq \Gammap$, where $A \simeq \lambda_\mathrm{p}^2$, $\lambda_\mathrm{p}$ and $\Gammap$ denote the polariton mode area, wavelength, and linewidth, respectively.
Since the polariton--polariton interaction strength ($\gpp$) is expected to be intrinsically weak, multiple approaches have previously been devised to ensure a combination of small $A$ and $\Gammap$ \cite{Schneider2017},  where sizable interactions within a polariton condensate were observed by measuring the interaction-induced polariton resonance energy blueshift \cite{Ferrier2011}.
The $\gpp$ extracted from these measurements can be significantly higher than the values expected from theory \cite{Ciuti1998}; comparable values were inferred from other measurements \cite{Mukherjee2019,Vladimirova2009,Brichkin2011}.
It is widely accepted that these large $\gpp$ values originate from the presence of a dark exciton reservoir created by the laser excitation \cite{Estrecho2019}.

Arguably, the true polariton--polariton interaction could be unequivocally determined through the degree of photon antibunching in correlation measurements. A promising avenue for observation of photon correlations is provided by open hemispherical cavities that allow for optical confinement of the mode area to a scale comparable to those obtained using micropillars \cite{Ferrier2011}: these structures show substantially prolonged polariton lifetimes due to the absence of losses on the pillar sidewalls and have the added benefit of in situ tunability of the cavity length \cite{Besga2015,Steinmetz2006,Hunger2010}.
Embedding an InGaAs quantum well (QW) in such an open cavity led to the first observation of finite quantum correlations between polaritons using pulsed excitation experiments \cite{Munoz-Matutano2019,Delteil2019}.
What is striking about these experiments is the unexpectedly large value of $\gpp$, extracted from the degree of photon antibunching. A possible explanation for the apparent discrepancy with theoretical predictions is the presence of a trion resonance due to charge accumulation in semiconductor structures that do not allow for control through electrical gates. Even though enhancement of $\gpp$ using the trion resonance is itself very interesting, the lack of charge control has adverse consequences: both experiments reported a relatively large $\Gammap$, which prohibited reaching the strongly correlated regime. Moreover, the large linewidth of the polariton mode made it impossible to carry out cavity--exciton and laser--polariton detuning-dependent photon correlation measurements using continuous-wave (cw) single-mode lasers that would have allowed for an accurate determination of $\gpp$.

We address this challenge by embedding an asymmetric InGaAs quantum well (QW) pair into a p--i--n heterostructure grown using molecular beam epitaxy with high-purity materials. 
Compared with previous designs, the addition of a p--i--n heterostructure not only allows for a precise control of the electric field at the QW position, but also reduces the likelihood of charge accumulation within the heterostructure by providing a drain channel.
While the asymmetric QW pair is designed to support indirect excitons, in this letter we investigate the system under flat-band conditions, i.e. where the indirect exciton is $\sim$\SI{12}{meV} higher in energy.
Additional measurements involving indirect excitons are presented in the supplement \cite{supp}.
To achieve maximal light-matter coupling, the QWs are positioned at the antinode of a linearly polarized TEM$_\mathrm{00}$ mode in a fiber-based hemispherical cavity with highly reflective DBR mirrors (\cref{sfig:Fig1-a}). We thereby mount the fiber on piezo-based nanopositioners, allowing for in situ tunability of the cavity length.
The curvature of the fiber facet provides the desired lateral mode confinement, reaching areas as small as $A = \SI{3.7}{\micro m^2}$ in our implementation.
\begin{figure}
    \includegraphics{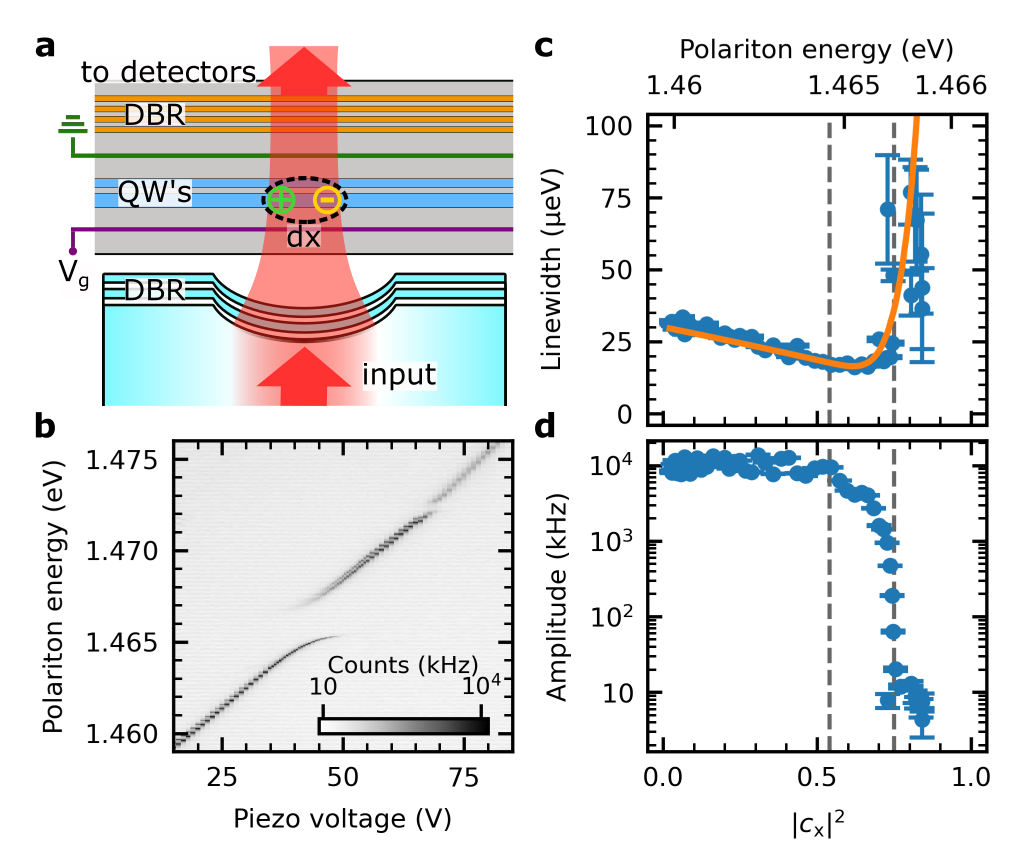}
    
    \caption{\justifying \textbf{a} Schematic of the hemispherical fiber cavity formed between a pair of DBR mirrors, enclosing the gate controlled \mbox{InGaAs} QW pair that hosts the direct excitons (dx). 
    \textbf{b} Polariton transmission spectrum as function of the piezo voltage used to tune the cavity length.
    It shows the lower- and upper polariton separated by the normal mode splitting ($2\Omega=\SI{3.003\pm0.006}{meV}$) and a smaller, additional anticrossing with a higher lying excitonic state.
   	\textbf{c} Linewidth and \textbf{d} peak transmission of the lower polariton resonance as function of exciton content and energy.
   	The fit of a model based on inhomogeneously broadened emitters coupled to a cavity \cite{Diniz2011} (orange line) to the measured linewidths allows to extract the exciton parameters $\gamma_\mathrm{inhom}=\SI{600\pm10}{\micro eV}$ and $\gamma_\mathrm{nonrad}=\SI{8.6\pm0.5}{\micro eV}$.
   	Vertical dashed lines in \textbf{c} and \textbf{d} indicate the detuning of the correlation measurement shown in \cref{fig:Fig2,fig:Fig3}.}
   	
    \label{fig:Fig1}
	\labelphantom{sfig:Fig1-a}
    \labelphantom{sfig:Fig1-b}
    \labelphantom{sfig:Fig1-c}
    \labelphantom{sfig:Fig1-d}
    
\end{figure}
In the experimental setup, the DBR coated fiber and the semiconductor sample are mounted upside down inside a liquid helium cryostat. 
Using the fiber for excitation and measuring the transmission through the cavity--QW structure from the polished backside of the substrate, we ensure that each detected photon originates from a polariton. 
A scan of the cavity mode across the exciton resonance reveals the two polariton branches in the spectrum (\cref{sfig:Fig1-b}), with a normal mode splitting of $2\Omega=\SI{3.003\pm0.006}{meV}$, characteristic of the strong coupling regime.
Due to the inherent birefringence of the GaAs substrate, the cavity eigenmodes are linearly polarized with a splitting of $\delta=\SI{250}{\micro eV}$. Since we observe $\Gammap<\delta$ (\cref{sfig:Fig1-c}), the lower polariton population can be excited selectively. Moreover, the energy of the lower polariton is lower than that of the dark exciton and trion resonances.
In the frame rotating with the excitation field, the corresponding Hamiltonian is given by
\begin{equation}
    H_\mathrm{p}=(\Delta-i\Gammap) p^\dagger p + \frac{\Upp}{2}p^\dagger p^\dagger pp + F^*p^\dagger +Fp,
    \label{eq:hamiltonian}
\end{equation}
where $\Delta=E_\mathrm{p}-E_\mathrm{l}$ is the detuning of the laser from the polariton resonance, and $F$ is the drive strength of the laser.
The polariton creation operator, $p^\dagger = c_\mathrm{c}^*a^\dagger+c_\mathrm{x}^*x^\dagger$, depends on the cavity--exciton detuning through the Hopfield coefficients $c_\mathrm{c}$ and $c_\mathrm{x}$; here, $a^\dagger$ and $x^\dagger$ denote the creation operators for the bare cavity and exciton modes, respectively.
The nonlinear term in the above Hamiltonian arises from the transformation of the exciton--exciton interaction into the polariton basis, leading to the generally presumed relation that $\gpp \propto |c_\mathrm{x}|^4$: as we detail below, however, this does not agree with our observations.

In \cref{sfig:Fig1-c} we show the linewidth of the lower polariton as a function of the cavity content. It exhibits a significant reduction by a factor of two (forty) compared to the bare cavity (exciton) linewidth. This is a direct result of the protection from the inhomogeneous broadening of the exciton ($\gamma_\mathrm{inhom}=\SI{600\pm10}{\micro eV}$) by the strong light matter coupling \cite{Kurucz2011,Diniz2011} in combination with a small non-radiative exciton linewidth ($\gamma_\mathrm{nonrad}=\SI{8.6\pm0.5}{\micro eV}$), which we attribute to the low density of impurities in our sample.  
The quoted exciton linewidths are obtained by fitting a model, based on the input-output formalism of inhomogeneously broadened emitters described in \cite{Diniz2011}, to the measured polariton linewidths (orange line in \cref{sfig:Fig1-c}).
This observation of exceptionally narrow polariton linewidths marks one of the key features of our system: since the corresponding polariton lifetime of $\sim\SI{50}{ps}$ exceeds the timing resolution achieved by state-of-the-art single photon detectors, we are able to fully resolve the time dependence of the polariton correlations.
Notably, the in situ tunability of the cavity length allows us to explore a wide range of polariton compositions, which is associated with significant changes of $\gpp$, $\Gammap$ and thus the resulting polariton--polariton correlations.
We are however limited to $\cx\leq\num{0.72}$, above which the overlap with the inhomogeneously broadened exciton distribution leads to an increase in linewidth and sharp drop in transmission.

We measure the polariton correlations as a function of both the cavity--exciton as well as the laser--polariton detunings by exciting a small population ($\langle p^{\dagger} p \rangle\le \num{0.05}$, estimated from the transmitted photon counts) of polaritons using a tunable, power-stabilized diode laser.
The correlations are computed from the waiting time distribution of photon arrival events, recorded by two \sspd's in a \hbt-like configuration combined with time-tagging electronics. The total timing uncertainty was measured to be $\sigma_\mathrm{tot}=\SI{14}{ps}$.
Further details on the data acquisition and analysis procedure are provided in the Supplemental Material \cite{supp}, including the additional measures required to mitigate mechanical drifts and fluctuations arising from the open-cavity design.
\begin{figure}
    \includegraphics{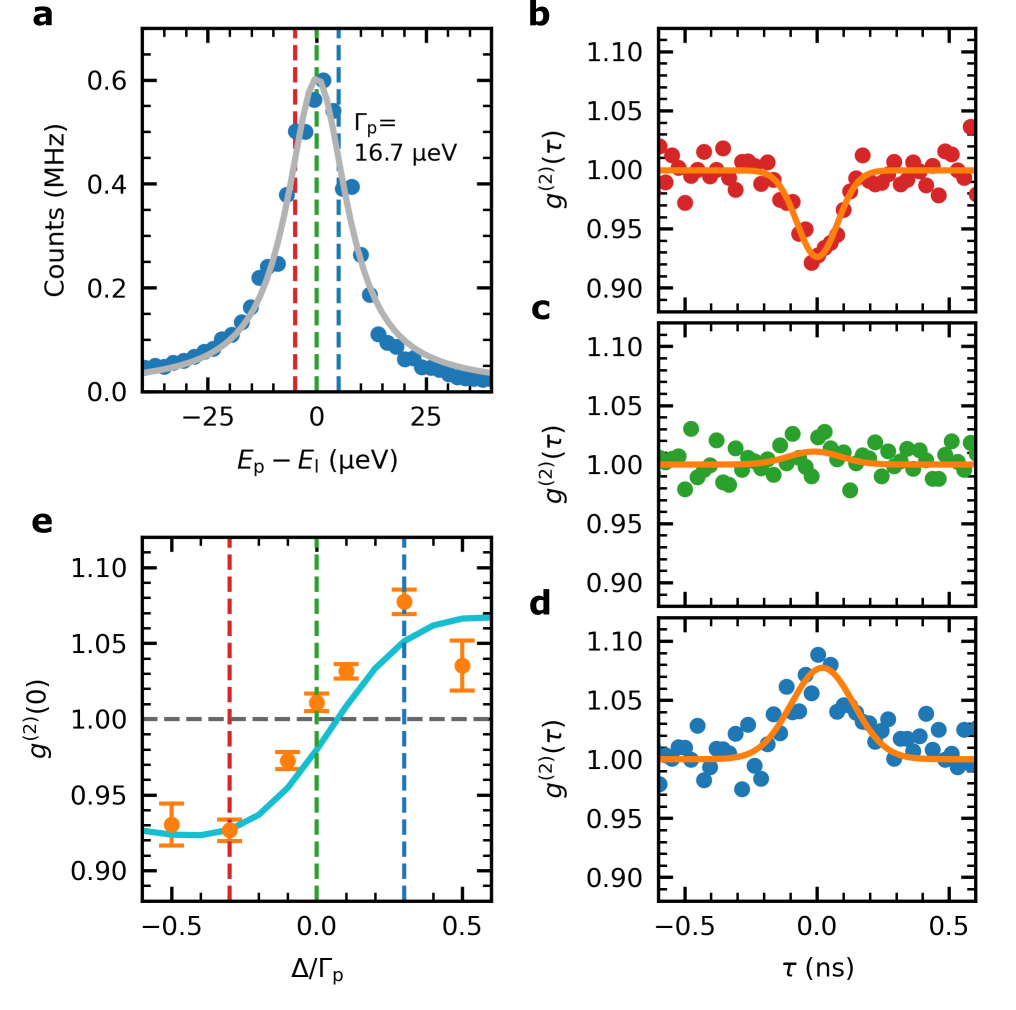}
    \caption{\justifying\textbf{a} Polariton transmission spectrum at an exciton content of $|c_\mathrm{x}|^2=0.72$ with a polariton linewidth of \SI{16.7}{\micro eV} extracted by a Lorentzian fit (gray line). 
    \textbf{b}, \textbf{c} and \textbf{d} $\gtwo(\tau)$ as function of the difference between photon arrival times ($\tau$) for the three detunings $\Delta=\qtylist{-0.3;0;0.3}{\Gammap}$.
    The observation of non-classical correlations for negative detunings, coherent correlations on resonance and bunching for positive detunings is a clear signature of the polariton--polariton interactions.
    A heuristic fit of a Gaussian function to the data (orange lines), allows the extraction of $g^{(2)}(0)$ as function of the detuning, shown in \textbf{e}.
    The numerical solution (turquoise line) to \cref{eq:hamiltonian}+\cref{eq:bxcoupling} with the parameters presented in the main text shows good agreement with the extracted values.
			}
	\label{fig:Fig2}
    \labelphantom{sfig:Fig2-a}
    \labelphantom{sfig:Fig2-b}
    \labelphantom{sfig:Fig2-c}
    \labelphantom{sfig:Fig2-d}
    \labelphantom{sfig:Fig2-e}
 
\end{figure}
The strongest correlations are expected from a combination of narrow linewidths and high excitonic fractions.
Indeed, when exciting polaritons with a negatively detuned laser at a cavity--exciton detuning corresponding to $\cx=\num{0.72}$ and a narrow linewidth of \SI{16.7}{\micro eV} (\cref{sfig:Fig2-a}), we observe a clear dip in the second order correlation function, as shown in \cref{sfig:Fig2-b}.
The observation of $\gtwo(0)<1$ is a strong signature of non-classical physics and is a direct consequence of the sizable polariton--polariton interactions.
We further show the correlations obtained when exciting the polaritons on resonance \cref{sfig:Fig2-c}, where the system retains the Poissonian statistics of the laser, and with a positive detuned laser, where we observed an enhanced two photon emission probability (bunching) in \cref{sfig:Fig2-d}.
The full dependence on the laser detuning is displayed in \cref{sfig:Fig2-e}, where we extract the values using a heuristic fit of a Gaussian function to the antibunching/bunching feature (see supplemental \cite{supp}).
It shows in more detail the transition from antibunching to bunching, following a "S"-shaped curve as expected from a system where $\gpp\lesssim\Gammap$ \cite{Verger2006}. We thereby find a maximal reduction of the two photon emission probability of \SI{8}{\percent} (see supplement for a more detailed analysis \cite{supp}).
\begin{figure}
    \includegraphics{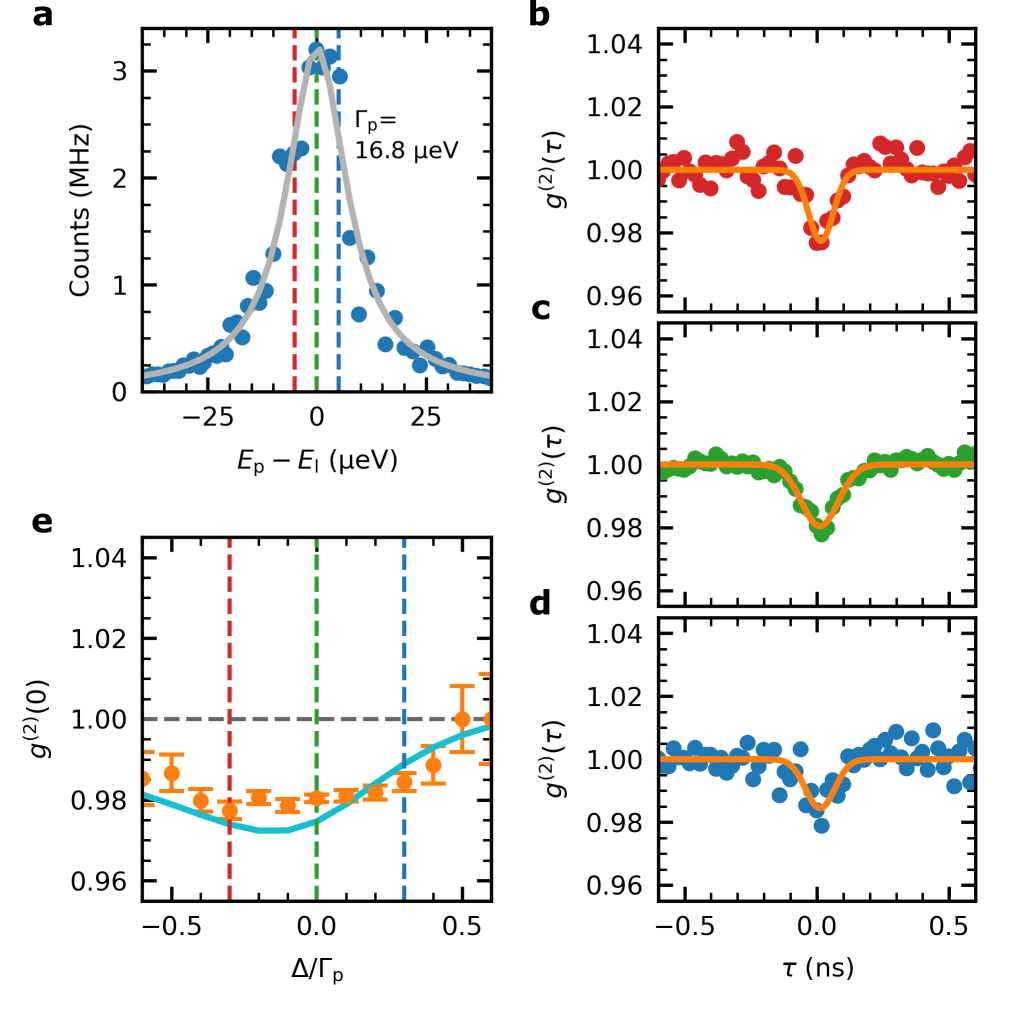}
    \caption{\justifying\textbf{a} Polariton transmission spectrum at an exciton content of $|c_\mathrm{x}|^2=0.54$ with a polariton linewidth of \SI{16.8}{\micro eV} extracted by a Lorentzian fit (gray line).
    \textbf{b}, \textbf{c} and \textbf{d} $\gtwo(\tau)$ as function of the difference between photon arrival times ($\tau$) for the three detunings $\Delta=\qtylist{-0.3;0;0.3}{\Gammap}$. A heuristic fit to a Gaussian function allows to extract $g^{(2)}(0)$ shown in \textbf{e}.
    Contrary to the expected "S" shaped dependence found for exciton like polaritons (\cref{fig:Fig2}), we observe a weak, detuning independent antibunching over a wide range. This behavior is well described by including the coupling to the biexciton, as demonstrated by the good agreement of the data with the numerical solution (turquoise line) of the Hamiltonian (\cref{eq:hamiltonian}+\cref{eq:bxcoupling}) with the parameters listed in the main text.}

    \label{fig:Fig3}
    \labelphantom{sfig:Fig3-a}
    \labelphantom{sfig:Fig3-b}
    \labelphantom{sfig:Fig3-c}
    \labelphantom{sfig:Fig3-d}
    \labelphantom{sfig:Fig3-e}

\end{figure}
When measuring at lower exciton contents we would simply expect the depth/height of the anti-/bunching to be reduced due to the weaker $\gpp$; our measurements however, exhibit a fundamentally different behavior. At $\cx=\num{0.5}$ (see \cref{sfig:Fig3-a}), we observe a small, detuning independent antibunching for $\Delta=\qtylist{-0.3;0;0.3}{\Gamma_p}$ shown in \cref{sfig:Fig3-b,sfig:Fig3-c,sfig:Fig3-d}, and the corresponding $\gtwo(0)$ over the full range of measured laser detunings in \cref{sfig:Fig3-e}. 
The widely used Hamiltonian given in \cref{eq:hamiltonian} fails to capture the observed correlations.

To explain this striking observation, we propose a dissipative/non-Hermitian blockade mechanism as described in \cite{Ben-Asher2023} based on two-polariton coupling to the biexciton resonance \cite{Carusotto2010}, described by the non-Hermitian Hamiltonian
\begin{equation}
\label{eq:bxcoupling}
    H_\mathrm{bx}=(\Delta_\mathrm{bx} - i\gamma_\mathrm{bx}/2) b^\dagger b + g_\mathrm{bx}\left(b^\dagger pp+bp^\dagger p^\dagger\right),
\end{equation}
where $b^\dagger$ is the biexciton creation operator, $\Delta_\mathrm{bx}$ is the two-photon detuning of the excitation laser from the biexciton resonance with binding energy $\varepsilon_\mathrm{bx}$, and $g_\mathrm{bx}$ is the biexciton--polariton coupling strength. We also introduce a phenomenological biexciton decay rate $\gamma_\mathrm{bx} \gg \Gamma_\mathrm{p}$.
The coupling $g_\mathrm{bx}$ together with the fast biexciton decay leads to excess line broadening of the two polariton state $\ket{2p}$, while the single-polariton state $\ket{p}$ remains unaffected.
This reduces their spectral overlap and, correspondingly, the two polariton excitation probability, as illustrated in \cref{sfig:Fig4-a}.
\begin{figure}
	\includegraphics{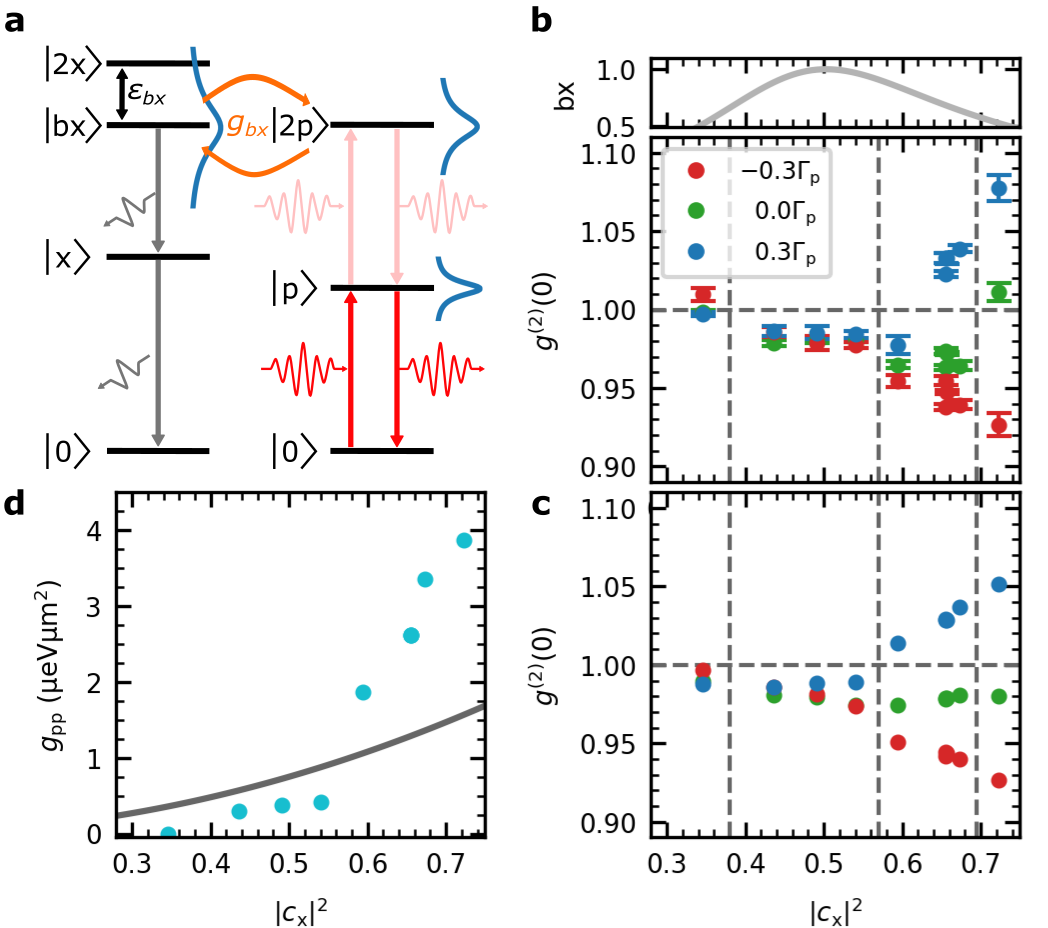}
	\caption{\justifying\textbf{a} Energy level scheme illustrating the dissipative blockade mechanism. 
		Due to the broadening of the doubly excited polariton state $\ket{2p}$, caused by its weak coupling to the biexciton $\ket{bx}$, the excitation probability of $\ket{2p}$, and consequently the two-photon emission probability, is substantially reduced.
		This results in antibunching over a wide range of cavity--exciton detunings, largely independent of the detuning between the laser and the polariton resonance.
		\textbf{b} $\gtwo(0)$ for three different laser detunings as a function of the exciton content compared to the estimated overlap of the biexciton (top panel).
		We distinguish different regimes depending on the detuning between the polariton and the biexciton. At high and low cavity contents the overlap with the biexciton is small and the correlations are determined by the polariton--polariton interactions or the coherence of the laser. Around the maximum of the distribution, the correlations are dominated by the dissipative coupling, followed by a hybrid regime at larger $\cx$.
		\textbf{c} The calculated values of $\gtwo(0)$ using the extracted parameters agrees well with the experimental data in $\textbf{b}$.
		\textbf{d} Polariton interaction strength ($\gpp$) as a function of $\cx$ estimated by matching the numerical solution of \cref{eq:hamiltonian}+\cref{eq:bxcoupling} to the correlation data.
		To highlight the deviation from the commonly used quadratic model, we also plot the interaction strength expected from the Born approximation \cite{Ciuti1998}.}
	\label{fig:Fig4}
	\labelphantom{sfig:Fig4-a}
	\labelphantom{sfig:Fig4-b}
	\labelphantom{sfig:Fig4-c}
	\labelphantom{sfig:Fig4-d}	
	
\end{figure}
In contrast to interaction-induced correlations, the antibunching depth resulting from this purely dissipative mechanism depends symmetrically on the detuning between the excitation laser and the polariton resonance, consistent with our observations in \cref{fig:Fig3}.
Furthermore, we expect the antibunching depth to depend to first order on the overlap with the biexciton, and therefore to be symmetric with respect to the detuning between the biexciton and the two-polariton state.
Our observations therefore differ from earlier reports \cite{Takemura2014,Scarpelli2024}, where a larger $g_\mathrm{bx}$ leads to a dependence expected from a non-dissipative Feshbach mechanism, manifested in a shift from repulsive to attractive interactions.
This symmetry in regard to the detuning between the resonance and the polariton is also where our observations differ from correlations that could be induced by a coupling to charges trapped in a local potential minimum. There the contribution to the correlations is expected to change asymmetric with respect to the detuning. 

In \cref{sfig:Fig4-b} we show $\gtwo(0)$ over a wide range of $\cx$ for three different laser detunings. The mechanisms dominating the observed correlations allows  to differentiate between four distinct regimes: Poissonian polariton correlations up to $\cx =\num{0.38}$; the Kerr-interaction dominated antibunching regime at $\cx=\num{0.72}$; the biexciton-resonance--dominated regime around $\cx=\num{0.5}$; and a crossover regime around $\cx=\num{0.65}$.
The range over which the biexciton coupling dominates the correlations, as well as the depth of the corresponding antibunching, is governed by the parameters $\varepsilon_\mathrm{bx}$, $\gamma_\mathrm{bx}$ and $g_\mathrm{bx}$, and is used to infer their values:
we find $\varepsilon_\mathrm{bx} = \SI{2.6}{meV}$, $\gamma_\mathrm{bx}=\SI{0.9}{meV}$ and $g_\mathrm{bx}=\SI{8}{\micro eV}$.
Based on these parameters, we can estimate the polariton--polariton interaction strength as function of the exciton content by matching the numerical solutions of $H_\mathrm{tot} = H_\mathrm{p}+H_\mathrm{bx}$ to the correlation data. 
The numerical results for $\gtwo(0)$ as function of $\Delta$ for $\cx=\qtylist{0.72;0.54}{}$ are plotted as turquoise lines in \cref{sfig:Fig2-e,sfig:Fig3-e}, and as function of $\cx$ in \cref{sfig:Fig4-c}, where they show good agreement with the experimental data.

The estimated values of $\gpp$ as function of the corresponding exciton contents are shown in \cref{sfig:Fig4-d}, clearly deviating from the commonly assumed quadratic dependence $\gpp=|c_\mathrm{x}|^4g_\mathrm{xx}$, added as gray line to the figure (using the exciton--exciton interaction strength obtained within the Born approximation $g_\mathrm{xx}=\SI{3}{\micro eV\micro m^2}$ \cite{Ciuti1998}). 
This discrepancy suggests that either our model based on the Hamiltonians \cref{eq:hamiltonian,eq:bxcoupling} or the quadratic scaling of the polariton interaction strength does not fully describe our system.
By restricting the discussion to large exciton contents, where the biexciton contribution is expected to be small, we find that the values we estimate are approximately a factor of $\sim\num{2.3}$ higher than expected from the Born approximation, which is comparable to the findings in \cite{Estrecho2019} at lower exciton contents. 
More importantly, our observations indicate that the quadratic scaling law originating from the Hopfield transformation does not capture the full details of the polariton interaction process. A similar observation has been made recently in theoretical predictions for dipolaritons in one and two dimensions \cite{Christensen2024,nakano2024}.

In this work we show how the reduced exciton non-radiative decay rate leads to a substantial reduction of the polariton linewidth, well below the linewidth of its constituents. This feature enables us to resolve the time-dependent correlations of the transmitted photons in a cw experiment over a wide range of cavity--exciton detunings and allows us to measure the dependence of $\gtwo(0)$ on the detuning of the laser from the polariton resonance. The maximum degree of photon antibunching of $\gtwo(0)=$\num{0.92\pm0.01} exceeds the values reported in prior experiments. Additional improvements in sample quality could further reduce the polariton linewidth, making it possible to achieve $\Upp>\Gamma_p$.
Moreover, we report a new regime of polariton blockade where quantum correlations are induced by a selective dissipation channel for the doubly excited polariton state, through the coupling to the biexciton resonance. 

\textbf{Acknowledgments} 
This work was supported by the Swiss National Science Foundation (SNSF) under Grant Number 200020$\_$207520.

\bibliographystyle{apsrev4-2}

%

\end{document}



\title{Supplemental Material: Quantum correlations and dissipative blockade of polaritons in a tunable fiber cavity}

\author{Gian-Marco Schnüriger}
\email{gm.schnueriger@uni-bonn.de}
\affiliation{Institut f\"ur Quantenelektronik, ETH Z\"urich, 8093 Z\"urich, Switzerland.}
\affiliation{Physikalisches Institut, Universit\"at Bonn, 53115 Bonn, Germany.}
\author{Martin Kroner}%
\affiliation{Institut f\"ur Quantenelektronik, ETH Z\"urich, 8093 Z\"urich, Switzerland.}
\author{Emre Togan}%
\affiliation{Institut f\"ur Quantenelektronik, ETH Z\"urich, 8093 Z\"urich, Switzerland.}
\author{Patrick Knüppel}%
\affiliation{Department of Physics, Universit\"at Basel, 4056 Basel, Switzerland.}
\author{Aymeric Delteil}%
\affiliation{Univerit\'{e} Paris-Saclay, UVSQ, CNRS, GEMaC, 78000 Versailles, France.}
\author{Stefan F\"alt}
\affiliation{Laboratorium für Festk\"orperphysik, ETH Z\"urich, 8093 Z\"urich, Switzerland.}
\author{Werner Wegscheider}
\affiliation{Laboratorium für Festk\"orperphysik, ETH Z\"urich, 8093 Z\"urich, Switzerland.}
\author{Atac Imamoglu}
\affiliation{Institut f\"ur Quantenelektronik, ETH Z\"urich, 8093 Z\"urich, Switzerland.}

\date{\today}

\maketitle

\section{Experimental setup}
\label{app:experimentalsetup}
\subsection{Sample and fiber}
The InGaAs quantum well (QW) sample used throughout this work was grown with molecular beam epitaxy. The structure shown in \cref{tab:singleqwstructure} is designed for optimal optical properties, by placing the doped layers at nodes of the cavity mode to minimize absorption while the QW pair is located in an antinode to maximize the light--matter coupling. To minimize the amount of current flowing through the structure when applying a static electric field between the doped layers we introduce AlGaAs/GaAs superlattices which act as tunnel barriers. The bottom distributed Bragg reflector (DBR) is designed for a center wavelength of \SI{850}{nm} formed by \num{24} layers of GaAs and \num{25} layers of AlAs, with an expected reflectivity of $R=\SI{99.992}{\percent}$.
\begin{table}
    \centering
    \begin{tabular}{lrll}
        Layer &Thickness (nm) &Material  & Repetitions\\
        \hline\hline
         Spacer& 40 & GaAs & \\
         p-doping& 40 & GaAs:C & \\
         Spacer&  99.7& GaAs & \\
         Tunnel block& 3 & AlGaAs & \\
         Spacer& 1 & GaAs & \\
         Tunnel block& 40 & AlGaAs & \\
         Spacer& 6.4 & GaAs & \\
         QW& 10 & InGaAs & \\
         QW separation& 11.8 & GaAs & \\
         QW& 4.6 & InGaAs & \\
         Spacer& 105.7 & GaAs & \\
         \hline
         Tunnel block& 7 & AlGaAs & \multirow{2}{*}{$\times$10}\\
         Tunnel block& 3 & GaAs & \\
         \hline
         Spacer& 23.2 & GaAs & \\
         n-doping& 30 & GaAs:Si & \\
         Spacer& 44.9 & GaAs & \\
         \hline
         DBR& 72.2 & AlAs & \multirow{2}{*}{$\times$34}\\
         DBR& 59.6 & GaAs & \\
         Buffer&  & GaAs & \\

    \end{tabular}
    \caption{Growth design of the sample.}
    \label{tab:singleqwstructure}
\end{table}

On the other hand, the curved surface on the fiber facet was fabricated by laser ablation with a highly focused CO$_2$ laser, creating a curved depression on the surface of the fiber tip \cite{Hunger2010}. The surface is then coated with \num{18}/\num{17} pairs of a Ta$_2$O$_5$/SiO$_2$ forming a (DBR) with an expected reflectivity of $R=\SI{99.99996}{\percent}$. The reflectance asymmetry between the substrate and the fiber mirror leads to \SI{98}{\percent} of the intracavity field leaking out through the semiconductor mirror, making it more efficient to detect photons from this side of the cavity. The fiber used in this work is the same as in \cite{Fink2018,Delteil2019} with a mode field diameter of \SI{2.18}{\micro \meter}.

\subsection{Excitation setup}
\begin{figure}
    \includegraphics{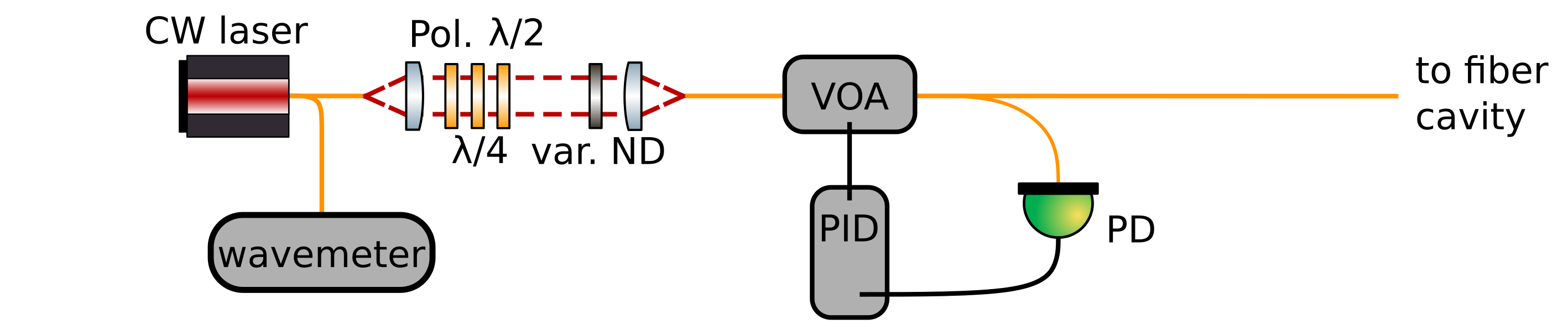}
    \caption{\justifying Setup for the excitation light source. The power-stabilization is done by a VOA controlled by a PID and to set the polarization we combine a polarizer (Pol.) with a half- and quarter-wave plate in a free space path. As the laser is used as energy reference for the experiment, a fraction of the light is sent to a wavemeter to obtain a precise readout of the wavelength.}
    	
    \label{fig:ExcitationSetup}
\end{figure}
We excite the cavity mode with a tunable, continuous wave diode laser (New Focus Velocity TLB-\num{6316}). 
The optical power is stabilized using a variable optical attenuator in combination with a PID controller, and we use a polarizer followed by a quarter- and a half waveplate to adjust the polarization, see \cref{fig:ExcitationSetup}. 
To accurately determine the wavelength of the laser we use a wavemeter (High Finesse WSU-30/661).

\subsection{Detection setup}
\begin{figure}
	\labelphantom{sfig:DetectorSetup-a}
	\labelphantom{sfig:DetectorSetup-b}
    \includegraphics{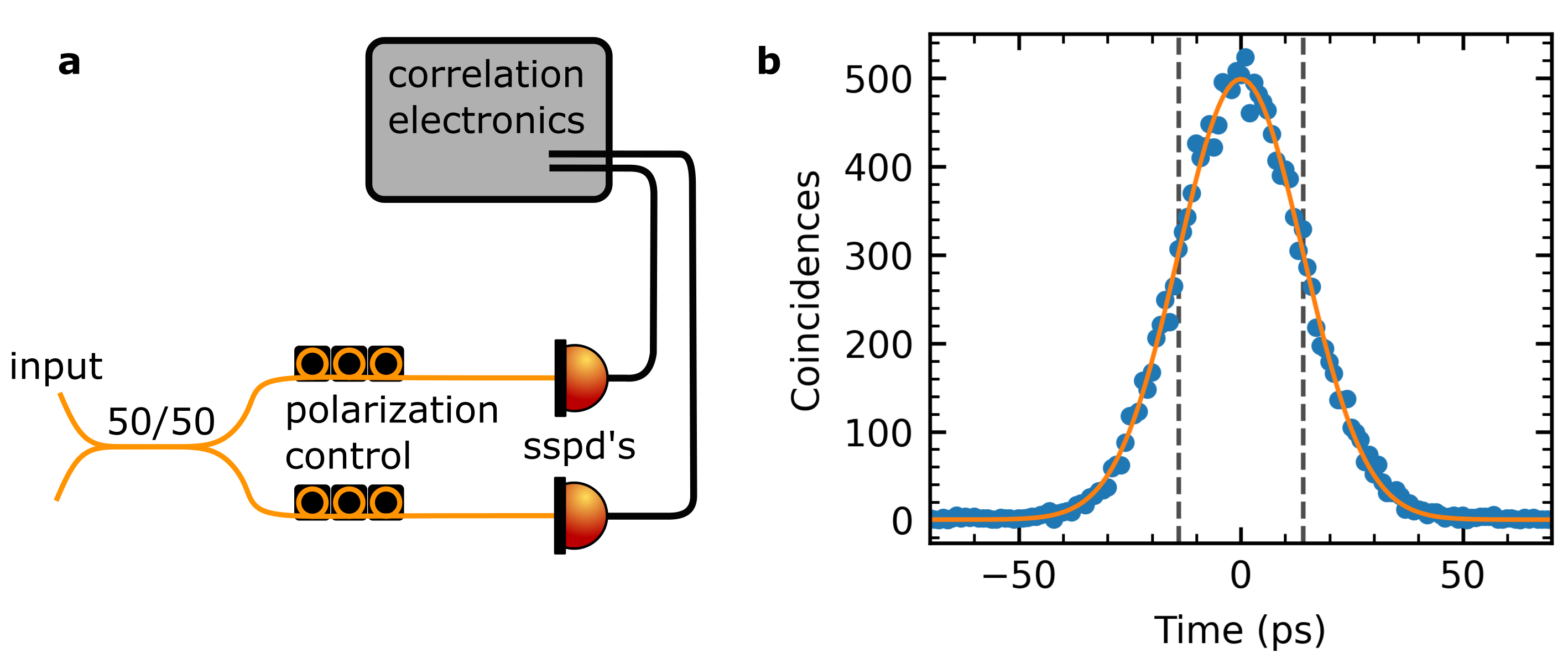}
    \caption{\justifying \textbf{a} Hanbury Brown and Twiss (HBT) like setup with detectors and time-correlated single photon counting electronics (TCSPC). The light coming from the sample is split with a fiber beam splitter and then sent to the two detectors. The arrival times of the electrical pulses from the two detectors are analyzed with tcscp electronics.
    \textbf{b} Autocorrelation of a \SI{1}{ps} long laser pulse. The detected pulse width is prolonged by timing uncertainty of the detector and the correlation electronics and is extracted by a Gaussian fit (orange line) $\sigma_\mathrm{tot}=\SI{14}{ps}$.
    }    	
    \label{fig:DetectorSetup}
\end{figure}
To resolve correlations of single photons on the timescales of the polariton lifetime $\sim\SI{50}{ps}$, we use a pair of superconducting nanowire single photon detectors (SSPD's) (Single Quantum, low jitter variant) in a Hanbury Brown and Twiss (HBT) like setup as illustrated in \cref{sfig:DetectorSetup-a}. 
The signal from the detectors is then read out and analyzed with time-correlated single photon counting electronics (TCSPC) (PicoQuant HydraHarp 400) by directly recording the arrival times of the pulses in the form of time tags.
To estimate the timing jitter of the combined correlation system we measure the autocorrelation of a \SI{1}{ps} long laser pulse shown in \cref{sfig:DetectorSetup-b}. The width of the signal is a sum of the jitter of the different components $\sigma_\mathrm{tot}^2=\sigma_\mathrm{elect}^2 + 2\sigma_\mathrm{det}^2 + 2\sigma_\mathrm{pulse}^2$ namely from the correlation electronics, the detector jitter and the duration of the laser pulse. Fitting a Gaussian curve to the signal allows us to extract the total jitter of $\sigma_\mathrm{tot}=\SI{14}{ps}$ and knowing the optical pulse length we can estimate the total jitter of the detection system to be $\sigma_\mathrm{sys}=\SI{13.9}{ps}$.

\section{Measurement procedure}
\label{app:Measurementprocedure}
\begin{figure}
    \includegraphics{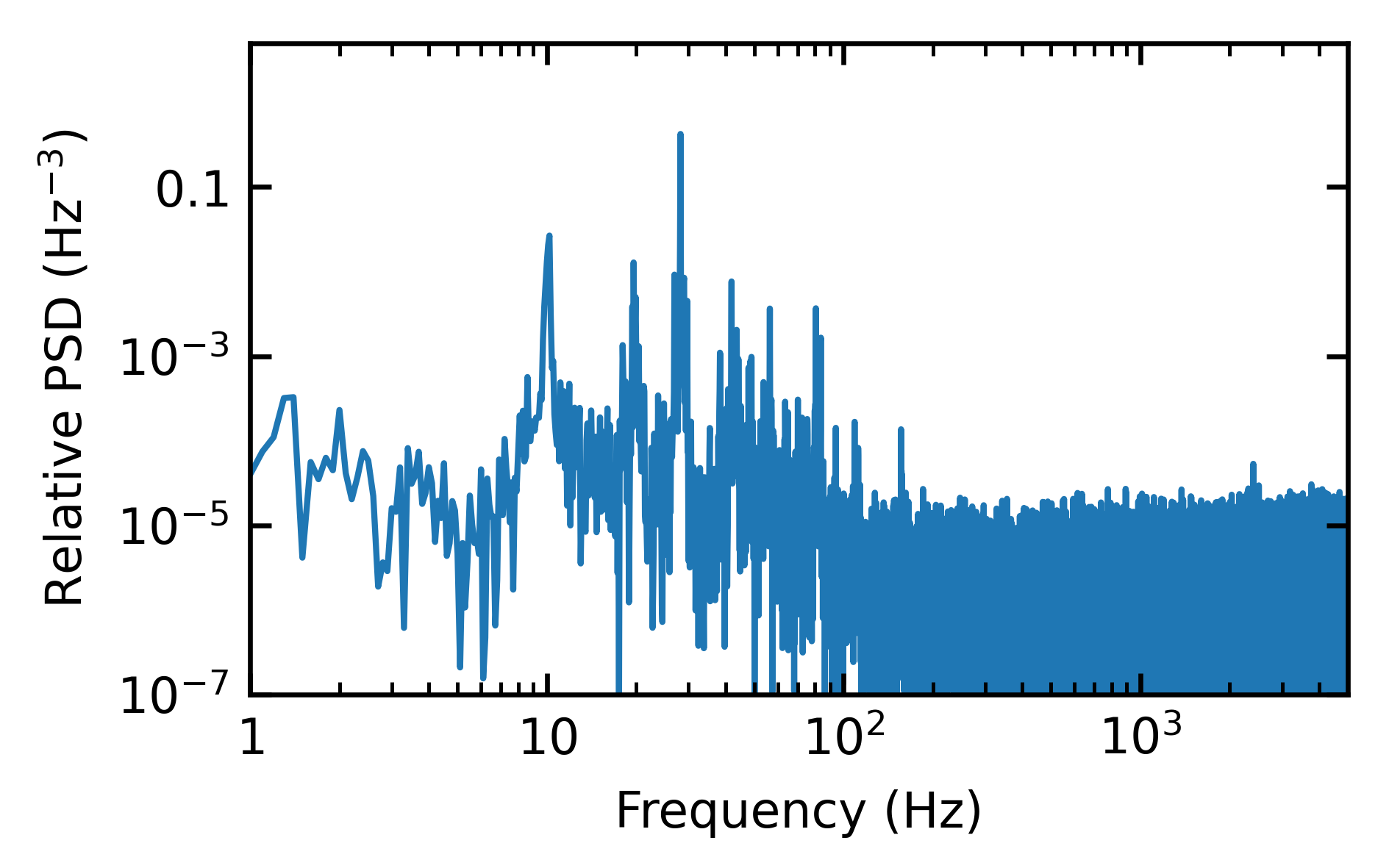}
    \caption{\justifying \textbf{a} Cavity noise spectrum obtained by Fourier transforming a normalized transmission time trace measured by exciting the cavity on the flank of the resonance. It shows that the dominant frequency components are situated below \SI{100}{Hz}.}    	
    \label{fig:CavityNoise}
\end{figure}
\begin{figure}
	\labelphantom{sfig:LaserModulation-a}
	\labelphantom{sfig:LaserModulation-b}
	\labelphantom{sfig:LaserModulation-c}
    \includegraphics{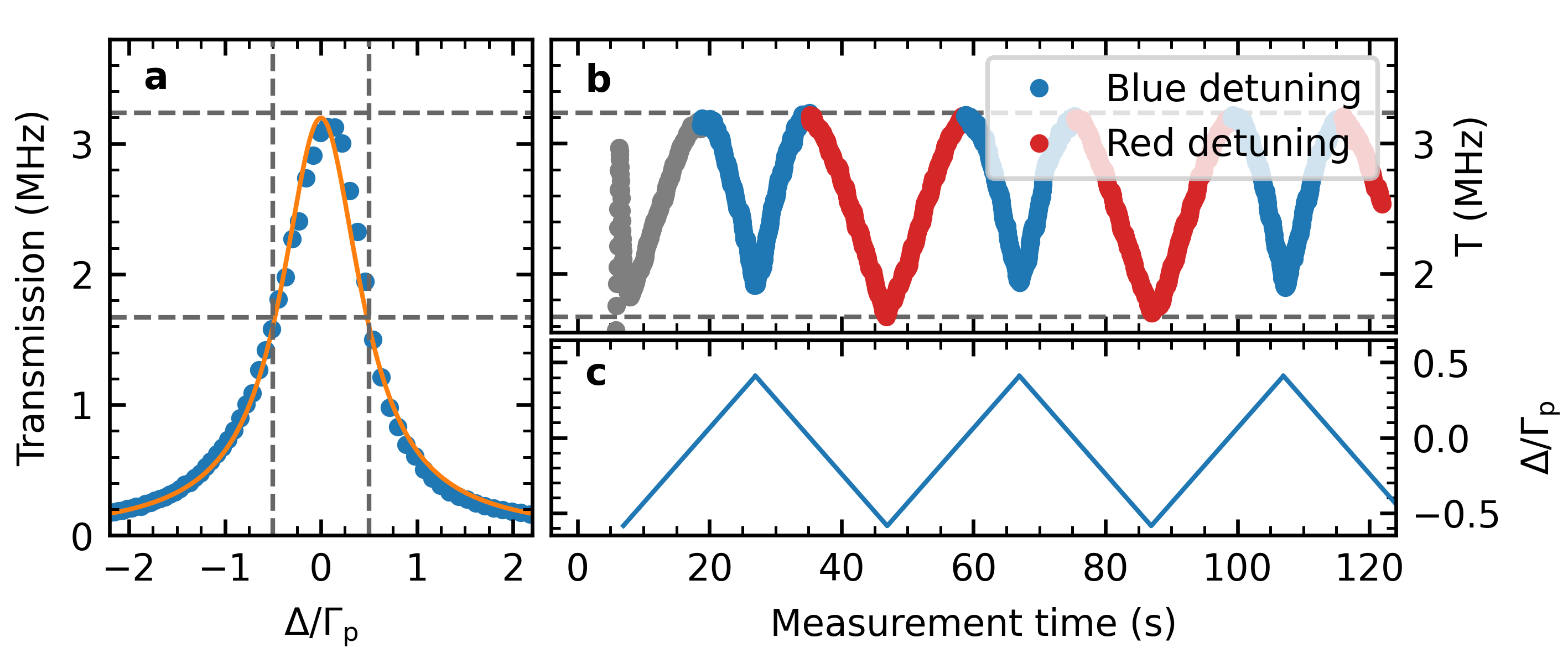}
    \caption{\justifying \textbf{a} Polariton mode together with the average countrate for each measurement chunk \textbf{b} modulated by sweeping the laser between \SIrange{-0.5}{0.5}{\Gammap} (\textbf{c}). By counting the maxima of the transmission we can assign each measurement chunk to positive or negative laser detunings $\Delta$.}    	
    \label{fig:LaserModulation}
\end{figure}

One of the main challenges in this system is to overcome the fluctuations of the cavity length introduced by the open cavity design. As can be seen in the noise spectrum in \cref{fig:CavityNoise}, these instabilities extend over a wide range of timescales and are governed on one end by slow drifts, such as creeping of the piezos and on the other end by acoustic vibrations that lead to fluctuations of the cavity length on timescales on the order of tens of milliseconds. Simultaneously the integration time of the correlation measurements necessary to obtain a high enough signal to noise spans from a few hours up to a day, due to the requirement for a low polariton density. 
\begin{figure}
	\labelphantom{sfig:PostSelection-a}
	\labelphantom{sfig:PostSelection-b}
    \includegraphics{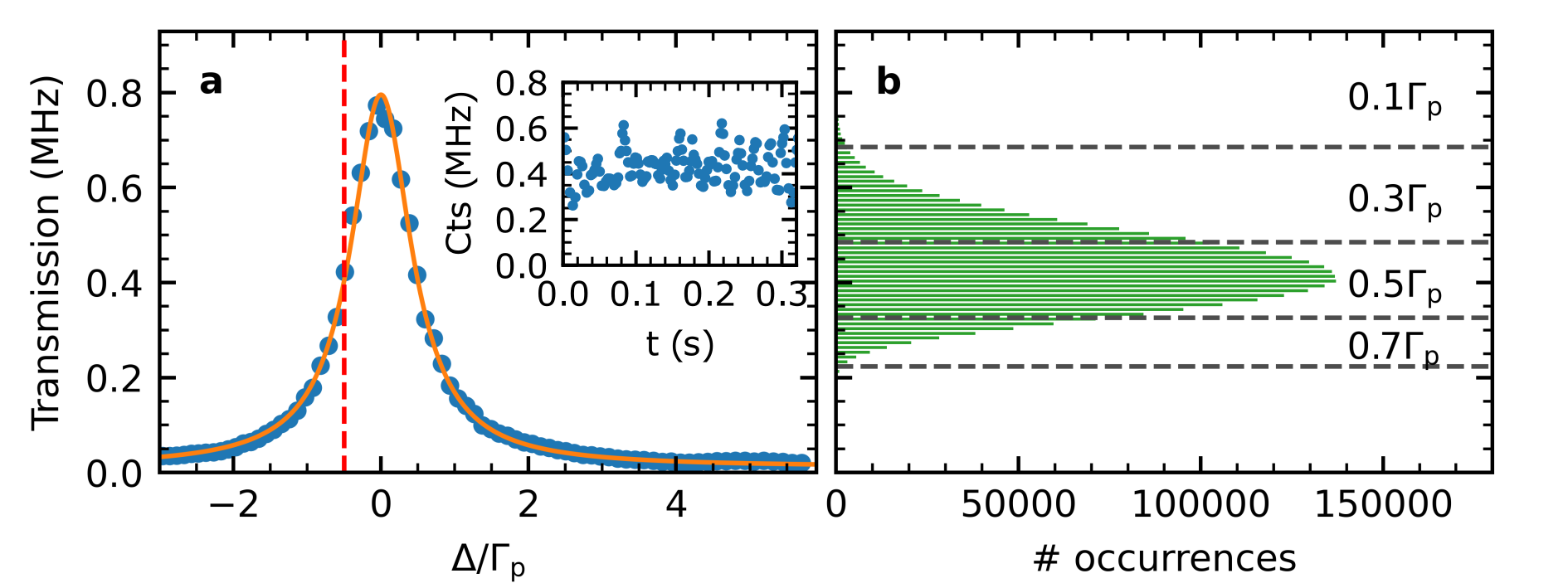}
    \caption{\justifying \textbf{a} Lorentzian line shape of the polariton and time trace of the countrate of each measurement chunk when exciting with a laser at $\Delta=\SI{-0.5}{\Gammap}$ (red line). \textbf{b} Histogram describing the number of measurement chunks occurring at a given countrate. Using the parameters obtained from the fit (orange curve) we can assign each of these chunks to a relative detuning $|\Delta|/\Gammap$, shown by horizontal lines for \numlist{0.1;0.3;0.5;0.7}.}    	
    \label{fig:PostSelection}
\end{figure}
To deal with these experimental challenges we combine different methods into our measurement procedure. First, we divide the whole measurement with a given set of parameters into multiple, repeatable intervals. In each repetition we first characterize the line shape of the mode to calculate the relative detuning $\Delta/\Gamma_\mathrm{p}$ between the excitation laser and the polariton and at the end of the iteration we sweep the gate voltage between \SIrange{-2}{2}{V} to flush out charges which might accumulate after prolonged exposure of the sample. 
To recover the detuning information from the measurement of the fluctuating cavity, we perform postselection of the timestamps. For this purpose we cut the sequence of time stamps obtained from the TCSPC electronics into shorter chunks and by calculating the countrate for each chunk we can assign the corresponding histogram to a detuning according to the countrate (\cref{fig:PostSelection}). Since we cannot distinguish between positive and negative detunings from the countrate alone, we slowly modulate the wavelength of the laser which allows us to track the changes in the detuning between the cavity and laser energy (\cref{fig:LaserModulation}). 

\section{Data analysis}
\subsection{Fitting procedure}
\label{app:Fittingprocedure}
Since the recovery time of the detectors is longer than the polariton timescales, we restrict our experiment to cross-correlations between the two detectors. The shown correlation data are then obtained by calculating the waiting time distribution of the photon arrival times. 

In principle, the true normalization of the coincidences does require the knowledge of the correlations up to arbitrary long timescales. But as we are only interested in timescales on the order of the polariton lifetime, we can treat data beyond \SI{1}{ns} as background. In \cref{sfig:g2_fit-a} we show raw data, with a clearly visible bunching background with a timescale of $\sim$\SI{19}{ns}. By removing a window of \SI{1}{ns} around zero time delay, show as gray area in the figure, we can quantify the background by fitting an exponentially decaying function, orange line, to the data. The results are then used to normalize the data and obtain $\gtwo(\tau)$ shown in \cref{sfig:g2_fit-b}.

In order to extract the magnitude of the correlation dip/peak ($\gtwo(0)$) independent of the underlying Hamiltonian, we fit a Gaussian to the data:
\begin{equation}
\label{eq:g2fit}
    \gtwo(\tau) = \left(1-\left(1-\gtwo(0)\right)e^{-\frac{(|\tau-\tau_0|)^2}{4\delta^2}}\right)
\end{equation}
where $\tau_0$ is the zero time delay of the detection setup and $\delta$ is the associated time scale proportional to the polariton lifetime. This model allows us to extract the value of $\gtwo(0)$ and its uncertainty considering the noise in the data. If there is no bunching or antibunching feature in the correlations, the fit tends to minimize the residuals by matching a small antibunching or bunching peak to points in the noise. These cases can be identified by their SNR ratio and large uncertainty in $\delta$ and since these values have no physical meaning, we set $\gtwo(0)=1$ with an uncertainty corresponding to the standard deviation of $\gtwo (\tau)$. 

\begin{figure}[h]
    \centering
    \labelphantom{sfig:g2_fit-a}
    \labelphantom{sfig:g2_fit-b}
    \includegraphics{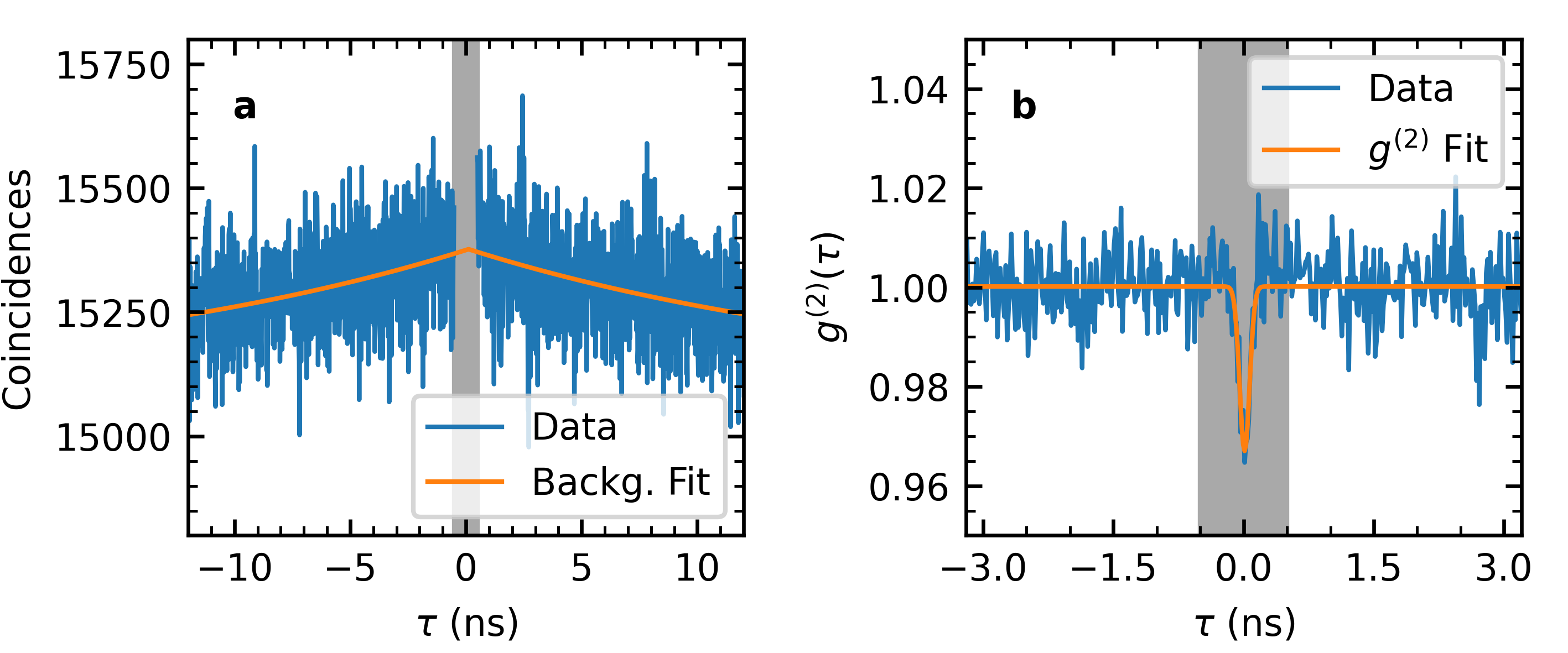}
    \caption[Correlation fitting procedure]{\justifying \textbf{a} Unnormalized second-order correlation function. The short timescales (gray area) are removed from the histogram of photon arrival time differences to fit an exponential decay (orange line) to the data. This is then used to obtain the normalized $\gtwo$ shown in \textbf{b}. By fitting \cref{eq:g2fit} to the data we obtain the depth and width of the dip, in this example: $\gtwo(0)=\num{0.953\pm0.005}$ and $\delta=\SI{51\pm8}{ps}$.}
    \label{fig:g2_fit}
\end{figure}

\subsection{Numerical calculations}
\label{app:nummcalc}
To compare the physical models to our data we perform numerical simulations based on the polariton Hamiltonian including the coupling to the biexciton \cite{Carusotto2010},
\begin{equation}
\begin{split}
\label{eq:hamiltonian_ap}
    H=&-\Delta p^\dagger p + \frac{U_\mathrm{pp}}{2}p^\dagger p^\dagger pp + F^*p^\dagger+Fp\\
    &- \Delta_\mathrm{bx}b^\dagger b + g_\mathrm{bx}\left(b^\dagger pp + bp^\dagger p^\dagger\right) ,
\end{split}
\end{equation}
together with the master equation
\begin{equation}
    \label{eq:masterequation_ap}
    \frac{\partial \rho}{\partial t}=\mathcal{L}\rho
\end{equation}
based on the corresponding Liouvillian \cite{Carmichael1999}
\begin{equation}
\begin{split}
    \mathcal{L} &= -\frac{i}{\hbar}\left[H,\rho\right]+\frac{\Gammap}{2} \left(2p\rho p^\dagger - p^\dagger p\rho - \rho p^\dagger p\right)\\
     &+\frac{\gamma_\mathrm{bx}}{2} \left(2b\rho b^\dagger - b^\dagger b\rho - \rho b^\dagger b\right).
\end{split}
\end{equation}
Where $p$, $b$ are the polariton/biexciton annihilation operators, $\Delta$/$\Delta_\mathrm{bx}$ are the detunings from the polariton/biexciton to the laser, $\Upp=\gpp/A$ is the polariton interaction energy, $F$ is the excitation amplitude in the polariton basis, and $g_\mathrm{bx}$ is the biexciton coupling strength. 
To solve these equations numerically we use QuTip \cite{Johansson2012,Johansson2013}, an open-source toolbox for python designed to simulate dynamical quantum systems. 

For the calculations we set the dimensions of the Hilbert space $N_\mathrm{p}\otimes N_\mathrm{bx}=5\otimes5$, and then use the build-in steady-state solver that calculates the Liouvillian using the corresponding collapse operators $\sqrt{0.5\,\Gammap}\,p$ and $\sqrt{0.5\,\gamma_\mathrm{bx}}\,b$, and solves for the steady-state density matrix $\rho_\mathrm{ss}$ satisfying $\mathcal{L}\rho_\mathrm{ss}=0$ using factorization. To then calculate the second-order correlation function
\begin{equation}
    \gtwo(\tau) = \frac{\expval{p^\dagger(0)p^\dagger(\tau)p(\tau)p(0)}}{\np^2},
\end{equation}
where $\np$ is the polariton number, the toolbox offers a master equation solver which integrates \cref{eq:masterequation_ap} for given times $\tau$ and then calculates the correlations. We can assume the polaritons to be in the steady state, therefore the initial state for the integration is $\rho_\mathrm{ss}$ and since we are interested in $\gtwo(\tau=0)$ we only integrate over short timescales to reduce the computation time. In the last step we normalize the correlations by $\np^2$ obtained by calculating the expectation value $\expval{\np}=\Tr{\rho_\mathrm{ss}p^\dagger p}$.
\section{Correlations at high exciton fraction}
At high exciton fractions the biexciton resonance is sufficiently detuned in energy to describe the correlation date with the bare, nonlinear polariton Hamiltonian:
\begin{equation}
    H=\Delta p^\dagger p + \frac{\Upp}{2}p^\dagger p^\dagger pp + F^*p^\dagger +Fp,
    \label{eq:hamiltonian}
\end{equation}
where $p^\dagger$ is the polariton creation operator, $\Delta=E_\mathrm{p}-E_\mathrm{l}$ is the detuning of the laser from the polariton resonance, $F$ is the laser drive strength, $\Upp$ the polariton interaction energy and $A$ the mode area. Using the numerical method described in \cref{app:nummcalc}, we can directly fit the correlation data shown in \cref{sfig:excitonlikecorrelations-b,sfig:excitonlikecorrelations-c,sfig:excitonlikecorrelations-d}, with $\Upp$ and $F$ as the free parameters. The best fit is drawn as brown line and shows good agreement with the correlation data and also with the detuning dependent values for $\gtwo(0)$ extracted by the Gaussian fit as described in \cref{app:Fittingprocedure} and shown in \cref{sfig:excitonlikecorrelations-e}. The thereby obtained interaction strength $\Upp=$\SI{4.3\pm0.2}{\micro eV \micro m\squared} is slightly higher than in the model incorporating the correlations induced by the biexciton, as there is still a finite overlap with the the biexciton which contributes to the depth of the antibunching. 

\begin{figure}[h]
    \centering
    \labelphantom{sfig:excitonlikecorrelations-a}
    \labelphantom{sfig:excitonlikecorrelations-b}
    \labelphantom{sfig:excitonlikecorrelations-c}
    \labelphantom{sfig:excitonlikecorrelations-d}
    \labelphantom{sfig:excitonlikecorrelations-e}
    \includegraphics{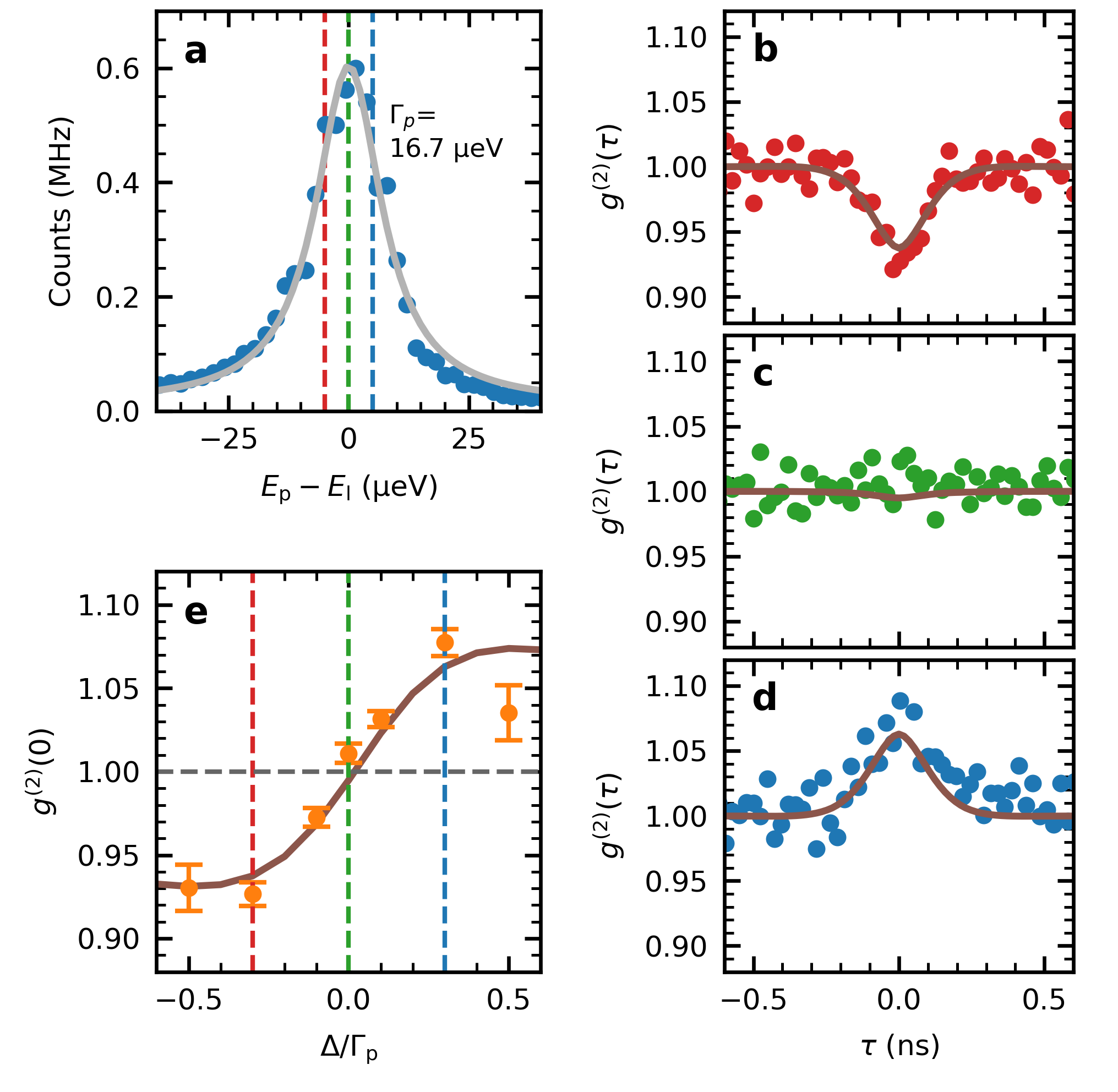}
    \caption{\justifying \textbf{a} Polariton transmission spectrum at a exciton content of $|c_\mathrm{x}|^2=0.72$ with a polariton linewidth of \SI{16.7}{\micro eV} extracted by a Lorentzian fit shown as gray line. 
    \textbf{b}, \textbf{c} and \textbf{d} second order correlation function depending on arrival time difference $\tau$ between for the three detunings $\Delta=\qtylist{-0.3;0;0.3}{\Gammap}$ indicated by vertical dashed lines in \textbf{a} and \textbf{e}. The brown line corresponds to a best fit to the bare, nonlinear polariton Hamiltonian \cref{eq:hamiltonian}.
    In \textbf{e} we compare the detuning dependence of $\gtwo(0)$ obtained by the heuristic Gaussian fit to the best fit to the full Hamiltonian.}
    \label{fig:excitonlikecorrelations}
\end{figure}

\section{Indirect exciton}
The asymmetric double QW structure allows us to introduce indirect excitons \cite{Cristofolini2012} in our system which we can tune into resonance with the direct exciton by changing the electric field as shown in \cref{sfig:ExcitonSpectrum-a,sfig:ExcitonSpectrum-b}. Due to the strong tunnel coupling both states hybridize into new eigenstates. And while the bare indirect exciton has a vanishing light matter coupling strength, the new eigenstates retain a fraction of the indirect exciton characteristics, while still coupling to light. Most importantly, the new state inherits a permanent dipole moment, leading to an  enhancement of the interactions in the system \cite{Byrnes2014,Nalitov2014}. The loss of oscillator strength at low gate voltages/high indirect exciton content can be seen in \cref{sfig:ExcitonSpectrum-c}, where we show the FWHM and peak area of the lower energy exciton mode. 
\begin{figure}
	\labelphantom{sfig:ExcitonSpectrum-a}
	\labelphantom{sfig:ExcitonSpectrum-b}
	\labelphantom{sfig:ExcitonSpectrum-c}
    \includegraphics{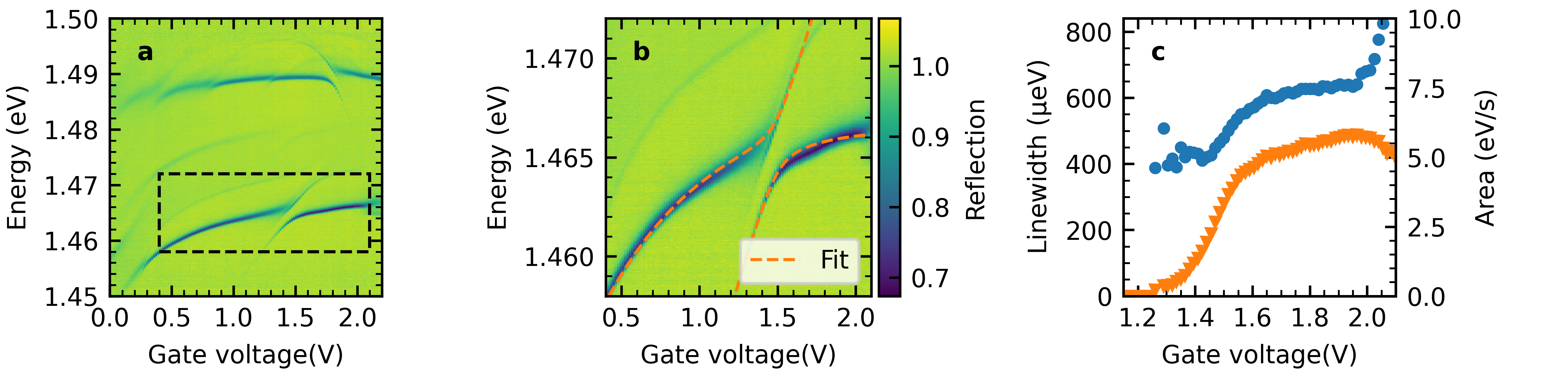}
    \caption{\justifying \textbf{a} Extended exciton reflection spectrum. The excitons from the two QW's shift quadratically with voltage due to the Stark effect. Each of them hybridizes with an indirect exciton which depends linearly on the voltage due to their dipolar nature. \textbf{b} Close-up of the lower anticrossing, together with a fit to the coupled oscillator model. \textbf{c} Linewidth (full width at half maximum) and area of the lower exciton branch. Both become smaller as the indirect exciton character increases.}    	
    \label{fig:ExcitonSpectrum}
\end{figure}

Coupling the hybridized exciton to cavity photons leads to three new eigenstates, the lower, middle and upper polariton. By applying a gate voltage and by changing the length of the cavity, we have control over the composition of the polaritons as shown in \cref{sfig:PolaritonSpectrum-a,sfig:PolaritonSpectrum-b,sfig:PolaritonSpectrum-c}.
\begin{figure}
	\labelphantom{sfig:PolaritonSpectrum-a}
	\labelphantom{sfig:PolaritonSpectrum-b}
	\labelphantom{sfig:PolaritonSpectrum-c}
    \includegraphics{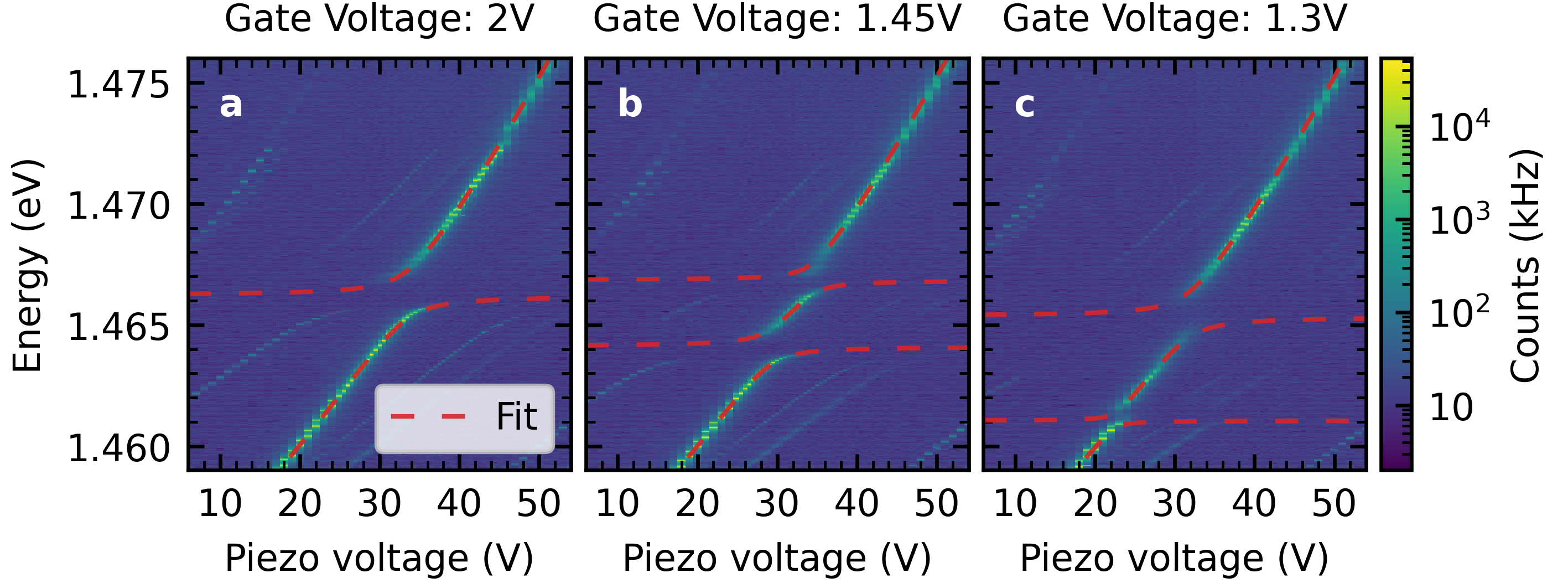}
    \caption{\justifying Polariton transmission spectrum as function of cavity length for different gate voltages. It shows the energy of the cavity mode scaling almost linearly with its length and the hybridization with the excitonic states. Going from left to right the indirect exciton is tuned past the direct exciton mode from being tuned to higher energies at \SI{2}{V}, to being on resonance at \SI{1.45}{V} to finally being slightly below the resonance at \SI{1.3}{V}. The red dashed lines correspond to a fit to a coupled oscillator model and show good agreement with the transmission data. }    	
    \label{fig:PolaritonSpectrum}
\end{figure}

\begin{figure}
	\labelphantom{sfig:Ix_linewidht_T-a}
	\labelphantom{sfig:Ix_linewidht_T-b}
    \includegraphics{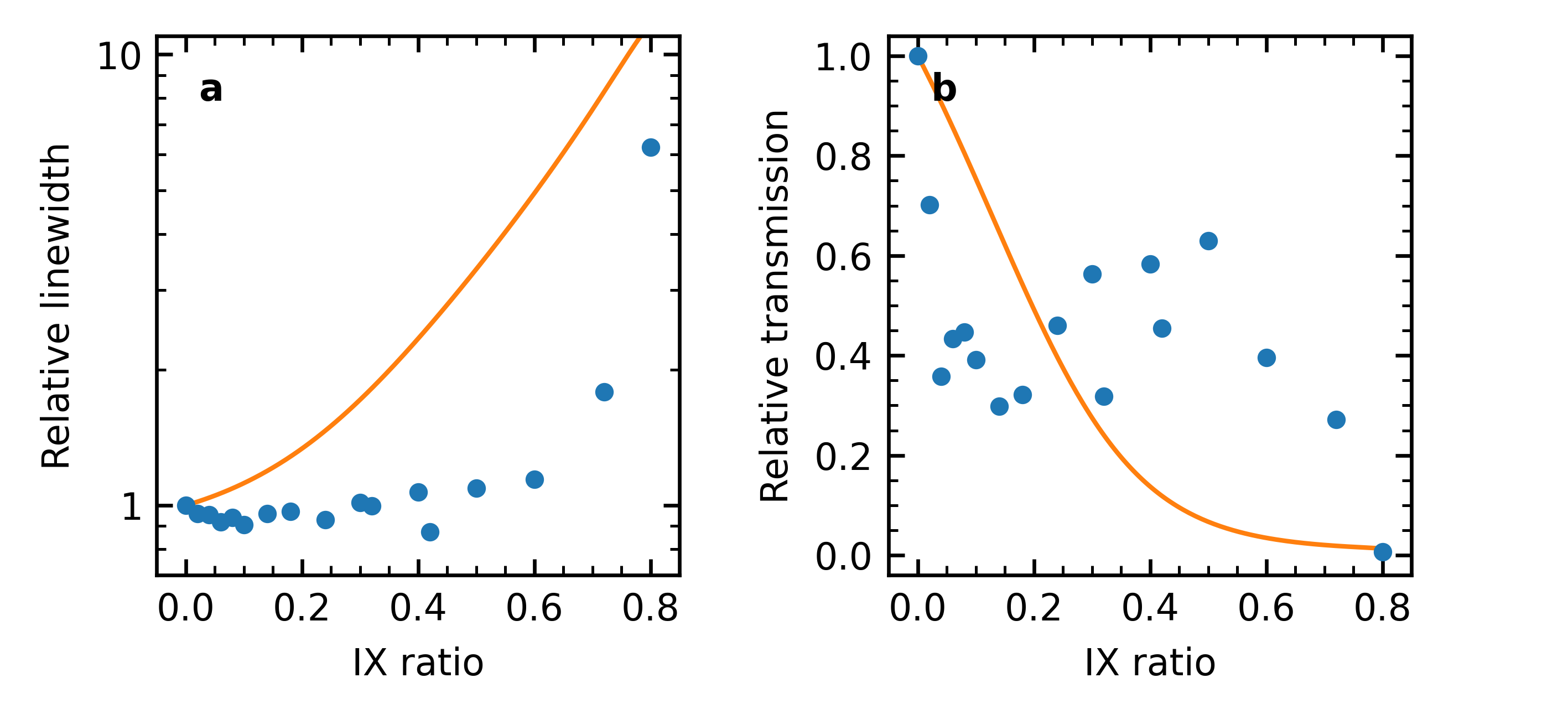}
    \caption{\justifying Relative change of the linewidth and transmission as function of the indirect exciton ratio at a constant cavity content $\cc=0.4$. The trend agrees with calculations assuming a cavity mode coupled to an inhomogeneously broadened lower exciton, with reduced light--matter coupling proportional to the indirect exciton ratio, $\Omega_\mathrm{eff}=(1-r_\mathrm{ix})\Omega$. This suggests that the loss of oscillator strength makes the polariton more susceptible to the inhomogeneous broadening of the emitter.}    	
    \label{fig:Ix_linewidht_T}
\end{figure}

While previous experiments have shown that we can expect a significant enhancement of the interactions by introducing indirect exciton content to the polariton \cite{Togan2018}, we do not observe this enhancement in the correlations. The reduction of the oscillator strength at higher indirect exciton contents introduces an upper bound on the mixture where measurements become unfeasible due to the low transmission through the cavity and the increased linewidth.
In \cref{fig:Ix_linewidht_T} we show the change in linewidth and transmission for a polariton with $\cc=\num0.4$ as a function of the indirect exciton ration $r_\mathrm{ix}=\cix/\left(\cdx+\cix\right)$.
While we observe an increase of the linewidth only at higher indirect exciton ratios, the transmission through the cavity drops severely, making correlation measurements increasingly difficult. The observed trend agrees with calculations (orange line \cref{sfig:Ix_linewidht_T-a,sfig:Ix_linewidht_T-b}) assuming a cavity mode coupled to the inhomogeneously broadened lower exciton branch, where the light--matter coupling depends on the indirect exciton ratio, $\Omega_\mathrm{eff}=(1-r_\mathrm{ix})\Omega$.

\begin{figure}
    \includegraphics{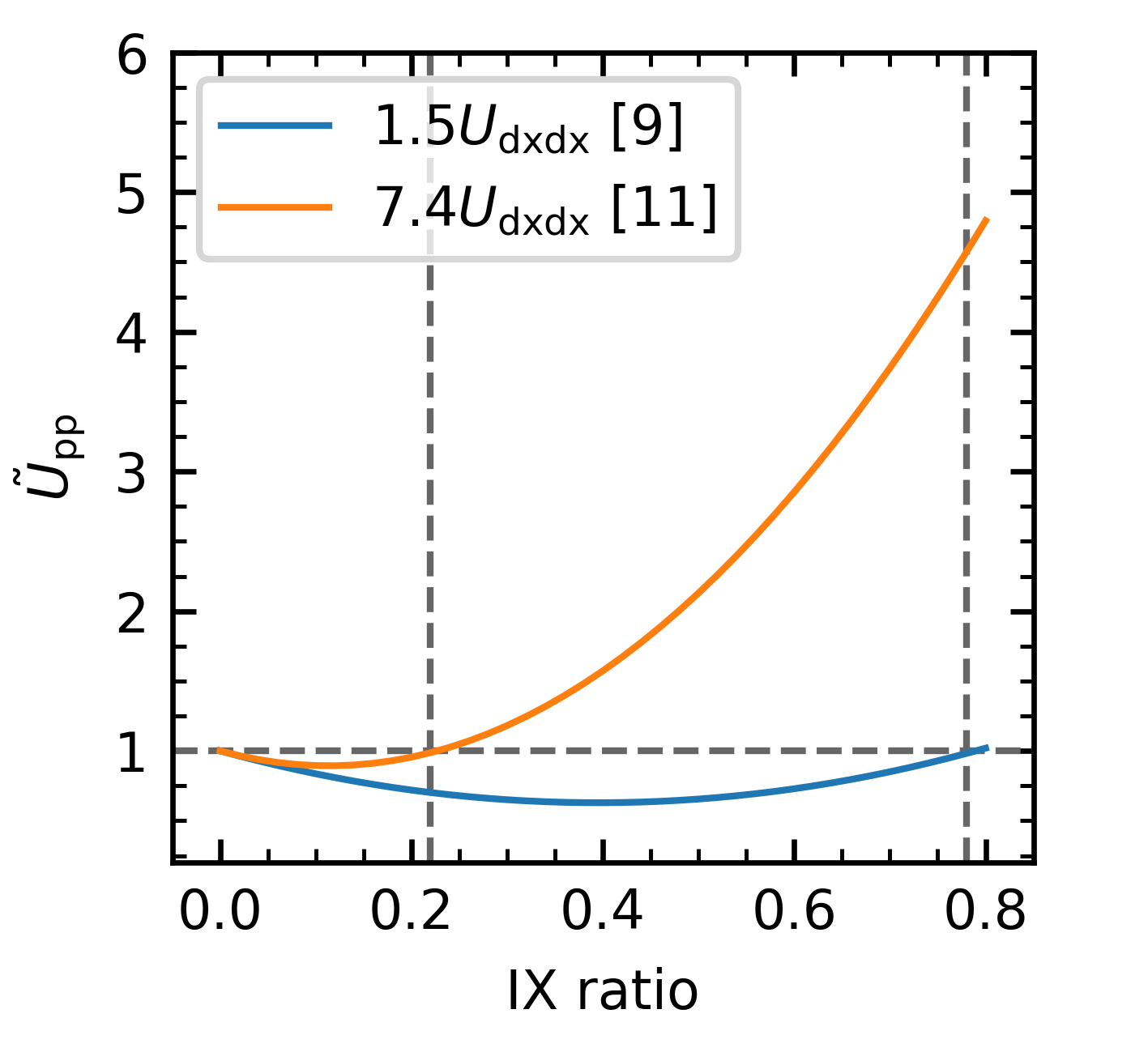}
    \caption{\justifying Enhancement of the normalized polariton interaction strength $\tilde{U}_\mathrm{pp}$ as a function of the indirect exciton ratio. Using values from theoretical predictions $U_\mathrm{ixix}=1.5\,U_\mathrm{dxdx}$ \cite{Byrnes2014}, and the findings from \cite{Togan2018} $U_\mathrm{ixix}=7.4\,U_\mathrm{dxdx}$, leads to different scalings but suggests that at a sizable indirect exciton ratio is necessary to enhance the interactions substantially.}    	
    \label{fig:Ix_Uppenhancement}
\end{figure}

By rewriting the dependency of the polariton interactions described in \cite{Byrnes2014} as 
\begin{equation}
    \tilde{U}_\mathrm{pp} = \frac{U_\mathrm{pp}}{U_\mathrm{dxdx}\left(1-\cc\right)^2} = \left(1-r_\mathrm{ix}\right)^2 + \frac{U_\mathrm{ixix}}{U_\mathrm{dxdx}}r_\mathrm{ix}^2,
\end{equation}
where we omitted the negligible interactions between direct- and indirect excitons, we can calculate the expected increase due to the indirect exciton. In \cref{fig:Ix_Uppenhancement} we show the values corresponding to $U_\mathrm{ixix}=1.5U_\mathrm{dxdx}$ as calculated in \cite{Byrnes2014} and for $U_\mathrm{ixix}=7.4U_\mathrm{dxdx}$ as found in \cite{Togan2018}. It shows that a significant indirect exciton content is required to enhance the interaction strength.

\begin{figure}
    \includegraphics{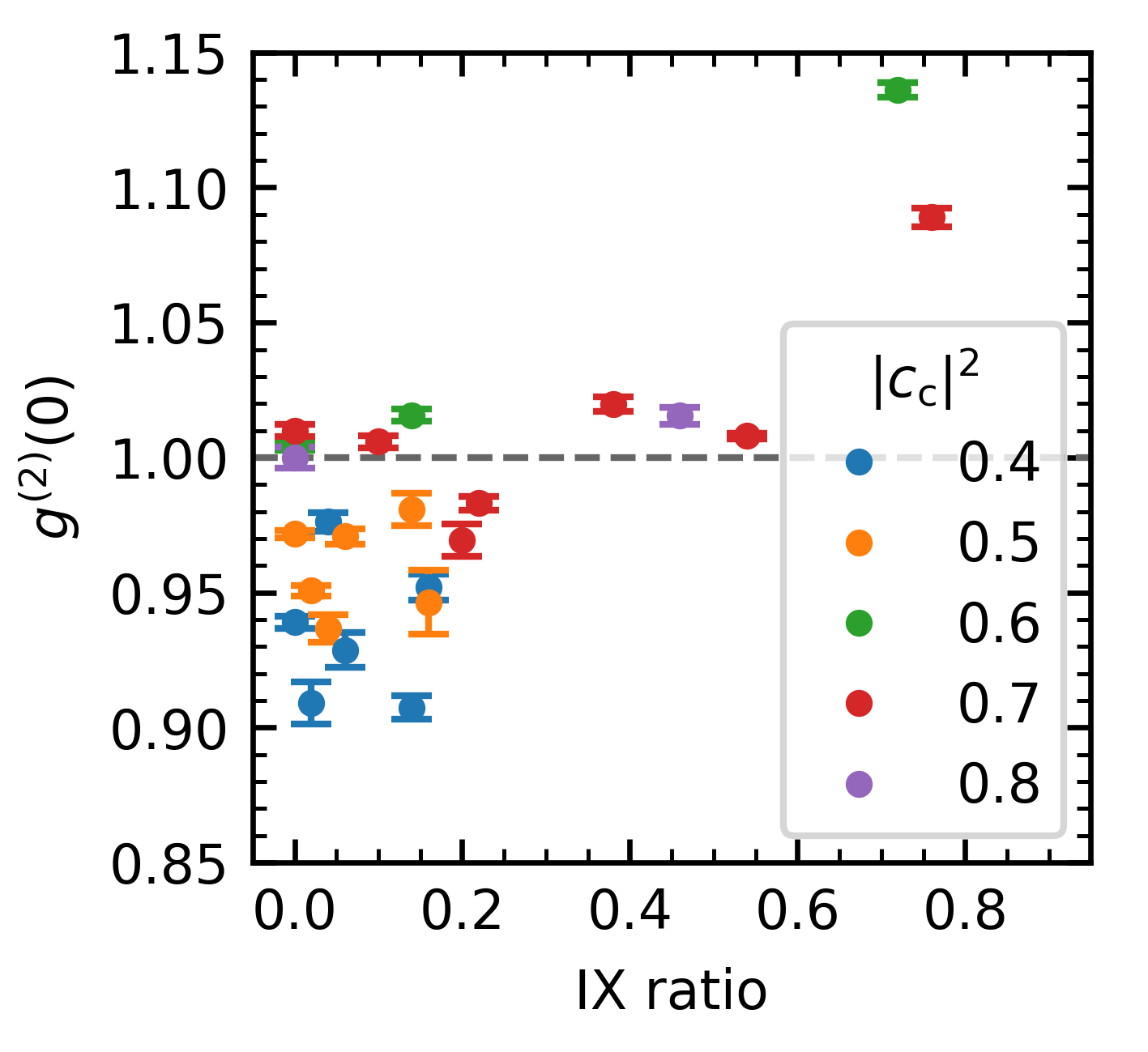}
    \caption{\justifying Summary of correlation measurements with different cavity contents as a function of the indirect exciton ratio. While we see nonclassical correlations in multiple measurements below \SI{30}{\percent} indirect exciton ratio, above that the narrow band in parameter space with low linewidth and sufficient transmission makes measurements difficult and we were unable to observe antibunching.}    	
    \label{fig:Ix_g2}
\end{figure}

In \cref{fig:Ix_g2} we show a representative overview of measurements done with finite indirect exciton ratios. For small ratios we do not expect an enhancement of the interactions but the transmission and linewidth are still favorable to perform measurements, resulting in multiple data points with nonclassical correlations. For higher ratios, where we expect an enhancement of the interactions, the low transmission only allowed for measurements with high cavity content and therefore no sizable antibunching was observed. The origin of the strong bunching features observed at high indirect exciton ratios in \cref{fig:Ix_g2} is not clear, and we cannot say if it is due to some spurious correlations from the sample or if it originates from the indirect exciton.

\bibliographystyle{apsrev4-2}
%